\documentclass{PoS}
\usepackage{epsfig}

\PoS{PoS(HEP2005)391}

\title{The first-level trigger of ATLAS}

\ShortTitle{The first level trigger of ATLAS}

\author{\speaker{Johannes Haller}
\thanks{on behalf of the ATLAS TDAQ LVL1 group: 
R.~Achenbach,
G.~Aielli,
A.~Aloisio,
M.~G.~Alviggi,
V.~Aprodu,
S.~Ask,
B.~M.~Barnett,
D.~Bartos,
B.~Bauss,
A.~Belkin
% ...
}

        CERN, Switzerland\\

        E-mail: \email{haller@mail.cern.ch}}

\abstract{
Due to the huge interaction rates and the tough experimental environment of pp collisions at a centre-of-mass energy $\sqrt{s}=14$\,TeV and luminosities of up to $10^{34}{\rm cm}^{-2}{\rm s}^{-1}$, one of the experimental challenges at the LHC is the triggering of interesting events.  In the ATLAS experiment a three-level trigger system is foreseen for this purpose. The first-level  trigger is implemented in custom hardware and has been designed to reduce the data rate from the initial bunch-crossing rate of 40\,MHz to around 75\,kHz. Its event selection is based on information from the calorimeters and dedicated muon detectors. This article gives an overview over the full first-level trigger system including the Calorimeter Trigger, the Muon Trigger and the Central Trigger Processor. In addition, recent results are reported that have been obtained from test-beam studies performed at CERN where the full first-level trigger chain was established successfully for the first time and used to trigger the read-out of up to nine ATLAS sub-detector systems.}

\FullConference{International Europhysics Conference on High Energy Physics\\

        July 21st - 27th 2005\\

        Lisboa, Portugal}

\begin{document}

\section{Introduction}
The LHC will collide protons at a centre-of-mass energy of 14 TeV with luminosities of up to $10^{34}{\rm cm}^{-2}{\rm s}^{-1}$. Bunches will cross with a rate of 40\,MHz, corresponding to a time interval between bunch-crossings (BC) of 25 ns. A total interaction (IA) rate of $\sim$\,1\,GHz is expected at nominal luminosity leading to $\sim$\,25 IAs per BC. In this challenging environment the ATLAS trigger system must reduce the rate to below the maximum rate that can be processed by the offline computing facilities, about 200\,Hz, while selecting previously undetected and rare physics processes. For example a Standard Model Higgs boson with a mass of 120\,GeV, decaying into two photons, is expected to occur in one out of $10^{13}$ IAs. The ATLAS trigger is composed of three levels. Its first level (LVL1) \cite{LVL1} is implemented in electronics and firmware, whereas the higher levels \cite{HLT} are based on software algorithms running in processor farms. In the following a brief overview of LVL1 is given and recent results from test-beam (TB) studies are reported. 

\section{LVL1 system overview}

\begin{figure}[t]
 	\begin{center}
\epsfig{file=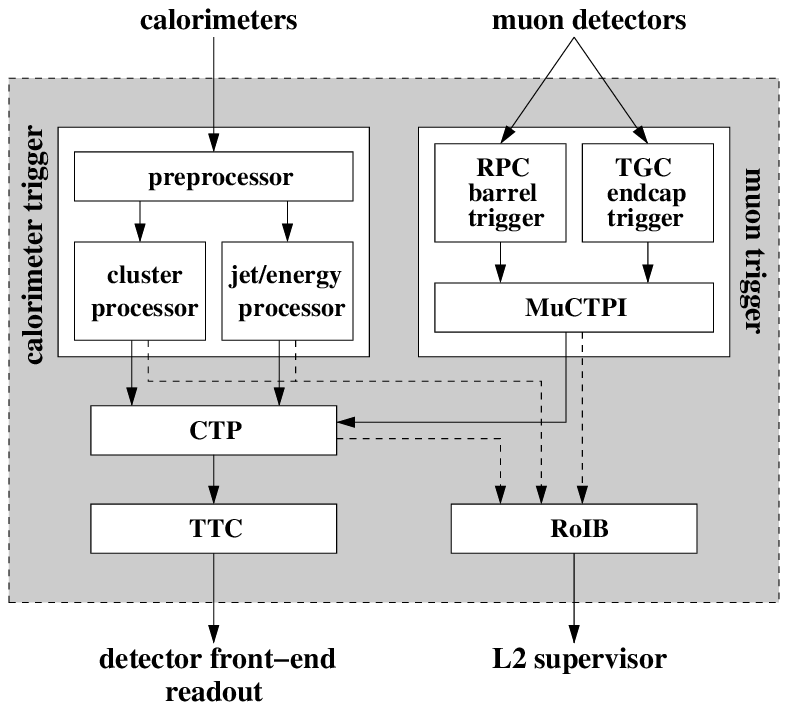,scale=0.8,angle=-90.}
\epsfig{file=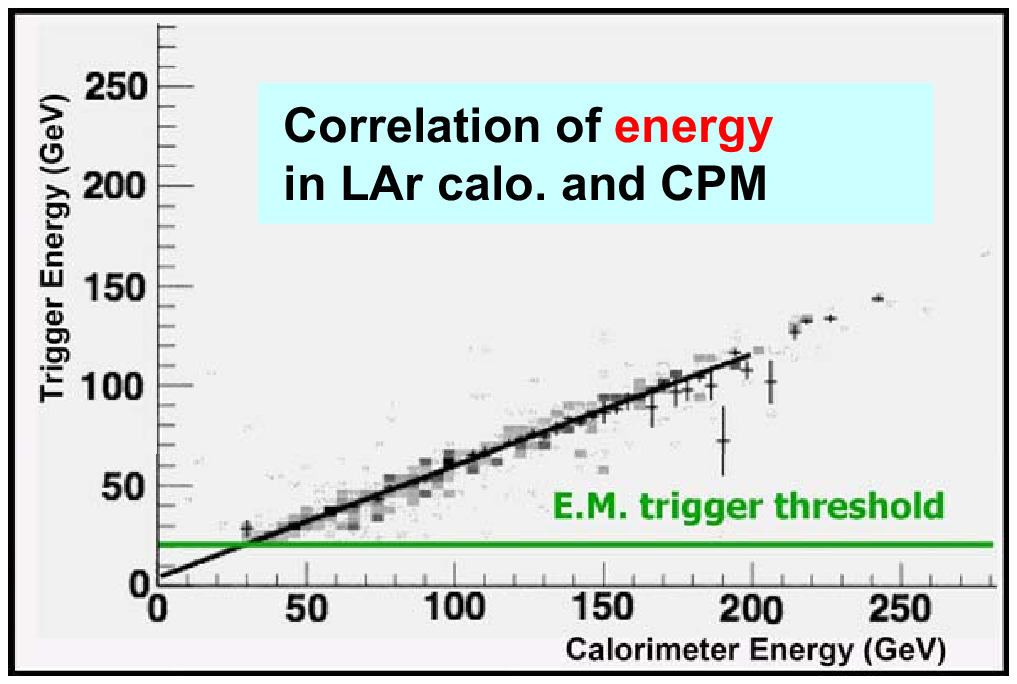,scale=0.6,angle=0.}
\vspace*{-5.8cm}
	\end{center}
	\caption{(left) Schematic view of the LVL1 trigger, (right) correlation of electromagnetic energy as measured by the LAr calorimeter read-out and the trigger electronics.}
	\label{fig:lvl1overview}
\end{figure}

The first-trigger level is a hardware-based system that reduces the event rate to below 75\,kHz (upgradeable to 100\,kHz) within a fixed latency of below 2.5\,$\mu$s.  
The LVL1 is composed of three parts (see Fig.~\ref{fig:lvl1overview} (left)): the Calorimeter Trigger (L1Calo), the Muon Trigger (L1Muon), and the LVL1 event-decision part implemented in the Central Trigger Processor (CTP).

The {\bf Calorimeter Trigger} relies heavily on FPGAs installed in the ATLAS electronics cavern. In ATLAS, calorimetry is provided by lead and copper liquid-argon sampling calorimeters (LAr) and an iron scintillator-tile sampling calorimeter (TileCal) for hadronic calorimetry in the barrel. On-detector electronics combines the analogue signals to $\sim$\,7200 projective trigger towers (TT). The Preprocessor (PPr) electronics  digitises the TT signals and performs BC identification and calibration.
The Cluster Processor (CP) identifies electron/$\gamma$ and $\tau$/hadron candidates using sliding window algorithms. The $E_{\rm T}$ of e/$\gamma$ ($\tau$/hadron) candidates is discriminated against up to 16 (8) programmable thresholds. 
The Jet/Energy Processor (JEP) identifies jet candidates and discriminates their $E_{\rm T}$ values against eight programmable thresholds. The JEP also evaluates several global energy sums. Synchronously with the 40\,MHz machine clock L1Calo sends multiplicities of e/$\gamma$, $\tau$/hadron and jet candidates, as well as the global energy information to the CTP via Common Merger Modules (CMM).

The ATLAS muon spectrometer consists of three stations of monitored drift-tube chambers (MDT) and dedicated fast muon detectors for triggering -- resistive-plate chambers (RPC) in the barrel and thin-gap chambers (TGC) in the forward region. The algorithms of the {\bf Muon Trigger} are based on hit coincidences in different stations within a geometrical road whose width is related to the $p_{\rm T}$ threshold applied exploiting the deflection of muons in the magnetic field. The coincidence logic allows six thresholds to be used at the same time. The Muon-to-CTP-Interface (MuCTPI) forwards the multiplicities of muon candidates for each threshold to the CTP after resolving possible double counting of muons that traverse more than one detector region. 

The {\bf Central Trigger} Processor makes the LVL1 decision (LVL1 accept, L1A) based on the information received from L1Muon, L1Calo and other sources (scintillator counters, random triggers, etc.). The CTP can handle up to 160 input trigger signals at any time and combines them logically to up to 256 triggers according to a trigger menu. It applies deadtime and prescale factors for each trigger. The L1A signal, the logical OR of all triggers, is then distributed to the various sub-detectors via Trigger Timing and Control (TTC) partitions including one Local Trigger Processor (LTP) each. A busy tree allows the sub-detectors to throttle the generation of L1As. For accepted events, all systems send data to the second-level trigger and to the read-out system via the Region-of-interest-Builder (RoIB).

\section{Results from studies at the ATLAS combined test-beam}
During 2004 a full slice of the ATLAS detector including the trigger detectors (LAr, TileCal, RPC, TGC) was installed at the H8 beam line of CERN's SPS providing p, e, $\pi$, $\mu$ and $\gamma$ beams with energies ranging from 1\,GeV up to 360\,GeV. The TB activity focused on testing prototypes and final modules of all sub-detectors, including the full trigger and data-acquisition chain. 
A data taking period with 25\,ns time structured beam offered the possibility to test the LVL1 trigger chain simulating LHC conditions with real detector signals~\cite{TESTBEAM}.

LAr and TileCal provided inputs to a full slice of L1Calo electronics (1\,\% of the final capacity), i.e. prototype modules with one PPr, one JEP, two CPs and teo CMMs. The modules successfully passed internal consistency checks and good correlation was found between the LAr and TileCal energy reconstruction and the energies as seen in the TT read-out from the trigger hardware. Figure~\ref{fig:lvl1overview} (right) shows the correlation of electromagnetic energy as measured by the LAr and the trigger electronics for a run with a trigger threshold of 20 GeV. The clear cut-off demonstrates that triggers were generated on genuine physics events. For the first time L1Calo was successfully integrated with the central trigger and provided trigger information to the CTP. 

\begin{figure}[t]

 	\begin{picture}(300,120)
	\put(20,0){	\epsfig{file=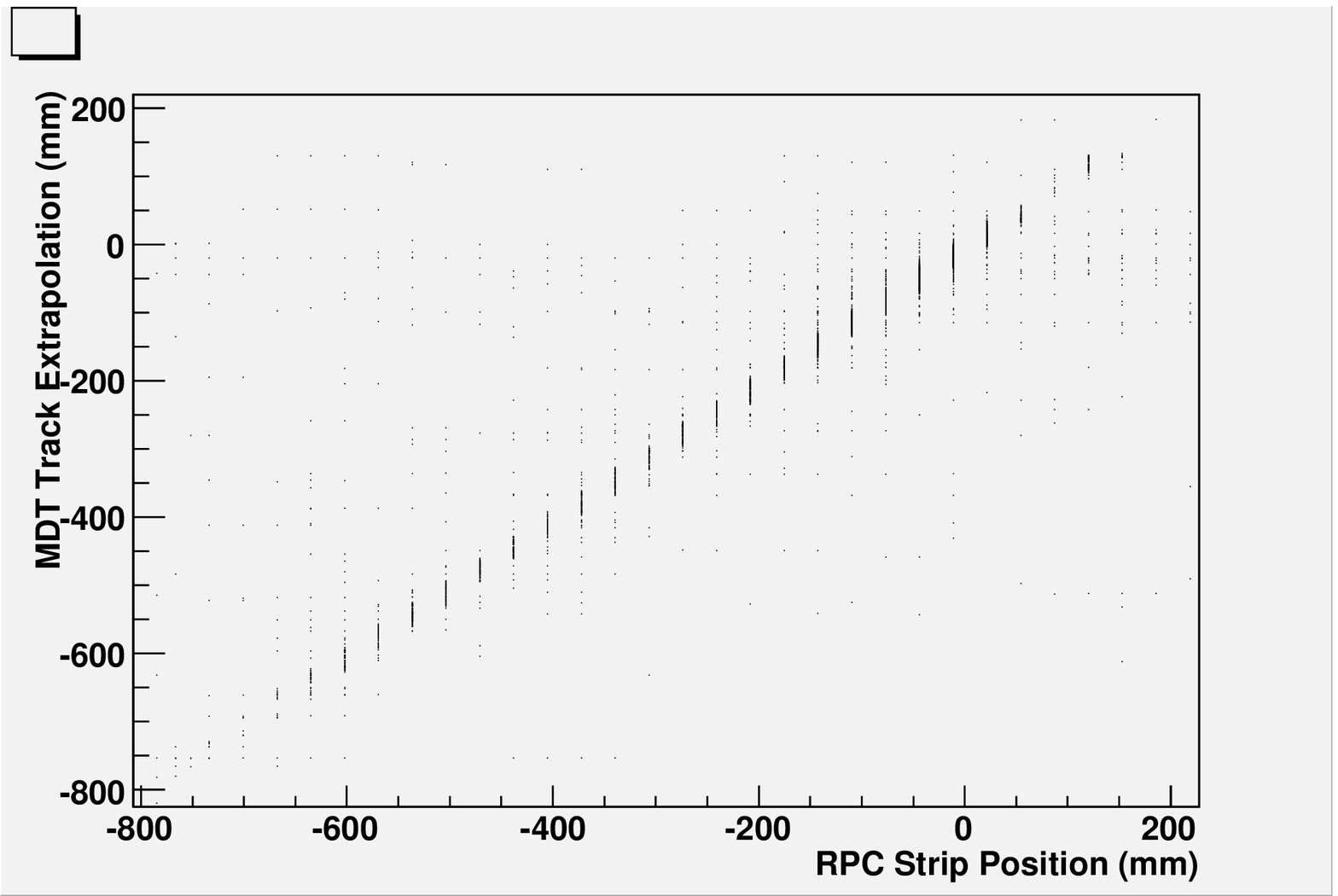,scale=0.37,angle=0.,clip=}}
	\put(250,138){	\epsfig{file=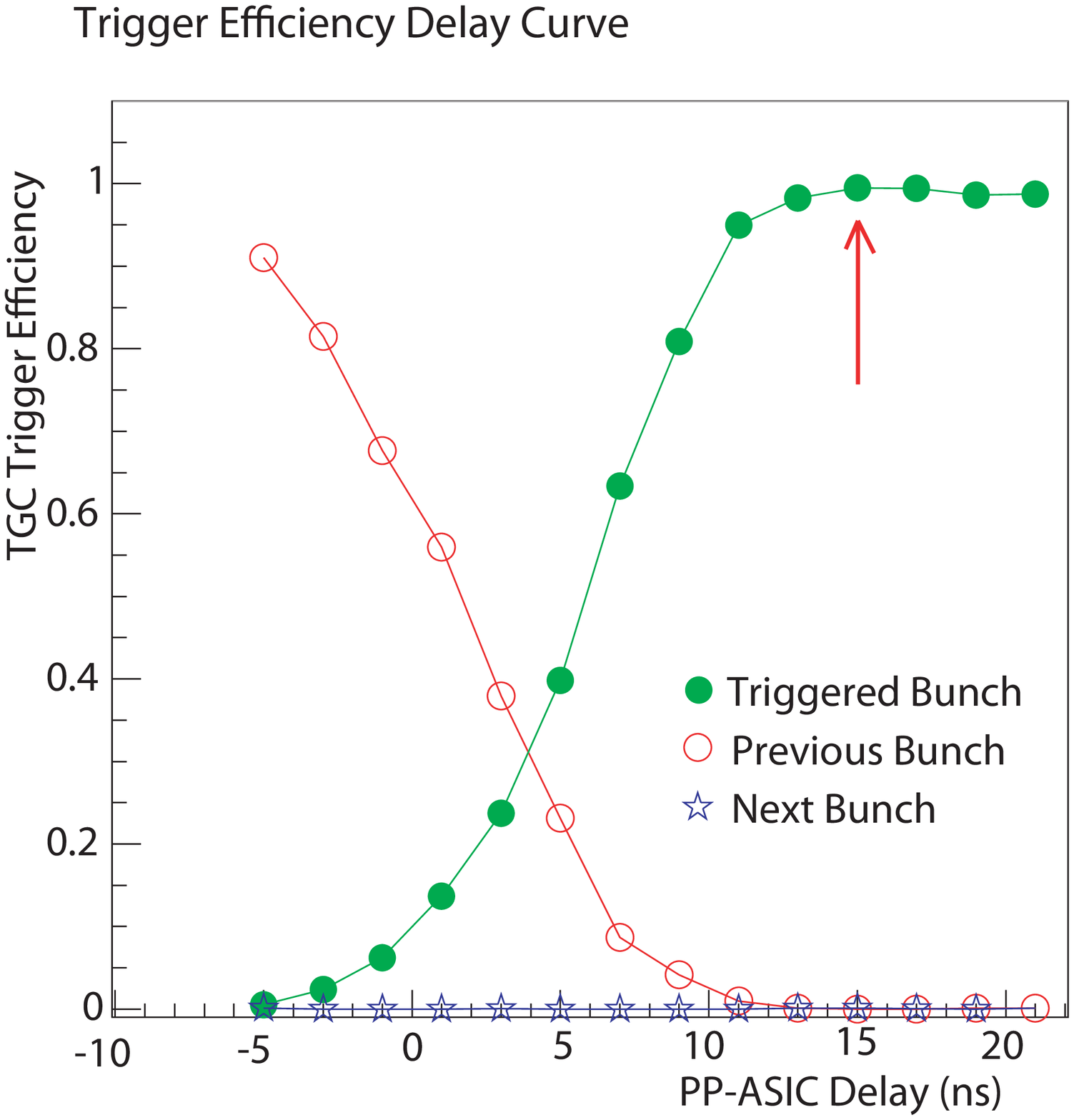,scale=0.25,angle=-90.}}
	\end{picture}
	\caption{Test-beam results: (left) correlation between RPC and MDT position measurements, (right) TGC trigger efficiencies as function of a delay parameter in the TGC electronics.}
	\label{fig:tb}
\end{figure}

Similarly, the L1Muon electronics was successfully integrated with the RPC and TGC detectors which delivered promising data. As an example Fig.~\ref{fig:tb} (left) shows the correlation between RPC and MDT position measurements. The efficiency to identify the correct BC by the endcap system is shown in Fig.~\ref{fig:tb} (right). It demonstrates the big timing margin in the TGC electronics where the efficiency to trigger on the correct bunch is large while the efficiency for the bunch before and after is tiny. Also the coincidence algorithm was successfully tested by emulation of the deflection in the magnetic field (missing at the TB) by shifting one of the trigger-chamber stations. The L1Muon electronics of both barrel and endcap was successfully integrated with the full LVL1 system and provided data to the MuCTPI that sent the multiplicities to the CTP. 

In total the final modules of the CTP included 46 input trigger bits (including 3 external bits from scintillator triggers of the beam instrumentation) in the calculation of the LVL1 decision. Using an LTP, the L1A signal was then fanned out to the sub-detector front-end electronics triggering their read-out. A projection of the measured LVL1 latency at the TB to the final ATLAS LVL1 latency including cable length and time-of-flight corrections gave a value of 2.13\,$\mu s$, which is well within the budget of 2.5\,$\mu s$. Also the busy tree was successfully established. The full LVL1 trigger system was run routinely under the main ATLAS Run Control system.

\section{Conclusion}
In this article a brief overview of the ATLAS LVL1 trigger has been given and promising results from recent TB studies have been reported. For the first time the full LVL1 trigger chain was established and integrated with the corresponding trigger detectors and the main ATLAS Run Control system. The L1A signal was used to trigger the read-out of up to nine sub-detectors.


\begin{thebibliography}{99}

\bibitem{LVL1} The ATLAS Coll., \emph{First-Level Trigger - Technical Design Report}, CERN/LHCC/98-14 (1998); R.~Spwioks {\it et al.}, \emph{The ATLAS Level-1 Central Trigger Processor (CTP)}, to app. in Proc. of LECC2005, Heidelberg; G.~Aielli {\it et al.}, \emph{The RPC LVL1 trigger system of the muon spectrometer of the ATLAS experiment at LHC}, IEEE TNS, 51 (2004) 1581; J.~Garvey {\it et al.}, \emph{The ATLAS level-1 calorimeter trigger architecture}, IEEE TNS, 51 (2004) 356.



\bibitem{HLT} The ATLAS Coll., \emph{HLT, Data Acquis. and Controls - Tech. Des. Rep,}, CERN/LHCC/2003-022 (2003).

\bibitem{TESTBEAM} R.~Achenbach {\it et al.}, \emph{Pre-Production Validation of the ATLAS Level-1 Calorimeter Trigger System}, to app. in Proc. of IEEE NPSS RT 2005, Stockholm; M.~Bianco {\it et al.}, \emph{Test Beam results and integration of the ATLAS Level-1 Muon Barrel Trigger}, IEEE NSS Conference Record, Vol. 1 (2004) 77; K.~Nagano {\it et al.}, \emph{Beam test of the ATLAS End-cap Muon Level 1 Trigger System}, ATL-DAQ-2004-010.  

\end{thebibliography}
\end{document}